# Multi-Modality and Temporal Analysis of Cervical Cancer Treatment Response

Haotian Feng
Emi Yoshida
Ke Sheng*
Dept. of Radiation Oncology
University of California-San Francisco
San Francisco, CA 94115
*Ke.Sheng@ucsf.edu

Cervical cancer presents a significant global health challenge, necessitating advanced diagnostic and prognostic approaches for effective treatment. This paper investigates the potential of employing multi-modal medical imaging at various treatment stages to enhance cervical cancer treatment outcomes prediction. We show that among Gray Level Co-occurrence Matrix (GLCM) features, contrast emerges as the most effective texture feature regarding prediction accuracy. Integration of multi-modal imaging and texture analysis offers a promising avenue for personalized and targeted interventions, as well as more effective management of cervical cancer. Moreover, there is potential to reduce the number of time measurements and modalities in future cervical cancer treatment. This research contributes to advancing the field of precision diagnostics by leveraging the information embedded in noninvasive medical images, contributing to improving prognostication and optimizing therapeutic strategies for individuals diagnosed with cervical cancer.

Keywords: cervical cancer, multimodality analysis, feature extraction, feature selection

## 1. Introduction

Cervical cancer is the fourth most common female malignancy worldwide and has been a significant global health burden. Cervical cancer is considered preventable, and early-stage detection is associated with significantly improved survival rates. Nevertheless, the disease remains a major cause of female mortality in low- and middle-income countries. In developed countries such as the US, approximately 13,820 new cases of invasive cervical cancer are diagnosed annually, and 4,360 women still die from the disease. Concurrent chemoradiotherapy is a mainstay treatment for cervical cancer. The treatment, however, is not personalized to individual patients despite a large observed variation in outcomes. Clinical stages, tumor histology, and positive lymph nodes were reported to be strong prognostic factors [1,2]. However, clinical factors alone are insufficient to explain differences in treatment response. Non-invasive medical imaging, particularly functional imaging modalities such as apparent diffusion coefficient (ADC) MR and positron emission tomography (PET), were used to study local tumor features. ADC quantifies the extracellular fluid compartment as a surrogate of cellularity. In highly proliferating tumors, high cellularity leads to lower ADC values due to restricted diffusion. Conversely, following chemoradiotherapy (CRT), necrosis may occur and result in increased ADC values as the tissue breaks down and water diffusion becomes less restricted. Nakamura et al. categorized patients into different groups and reported that the mean

ADC value predicts disease recurrence by analyzing the ROC curve [3]. Harry et al. found a significant correlation between early treatment ADC values and the response, as well as between percentage change in ADC and treatment response [4]. Similar results were reported by other investigators [5,6]. $^{18}$F-FDG PET/CT measures glycolysis, which is elevated in actively metabolizing tumor cells, and is a sensitive measure for the detection of nodal or distant metastases in cervical cancer [7]. A high tumor baseline FDG uptake has been correlated with worse outcomes [8]. Besides basic imaging features, such as standardized uptake value (SUV) and tumor volume, quantitative imaging analysis extracting intricate first, second, and third-order imaging textures was developed for medical images. Lucia et al. showed that radiomics features based on DWI-MRI and PET are independent predictors of recurrence and loco-regional control with significantly higher prognostic power than clinical parameters alone [9]. The same researchers further validated that high prediction accuracy of disease-free survival (DFS) and locoregional control (LRC) can be achieved with the combination of MR and PET radiomic features [10]. Zhao et al. demonstrated that radiomic analysis of MR first order and texture features could distinguish the early stages of cervical cancer and further assist clinical diagnosis [11].

Besides image analyses at a single pretreatment time point, researchers studied longitudinal changes with CRT. Salvo et al. used MRI to quantify tumor size changes [14]. Huang et al. examined three sequential DCE-MRI scans and showed that the number of low DCE voxels significantly correlated with cervical cancer treatment outcomes [15]. The same authors further established a universal threshold to quantify at-risk tumor voxels [16]. Bowen et al. proposed an image histogram intensity-based approach to quantify the tumor heterogeneity for individualized therapy decision support. The authors observed significant statistical changes in different modality images during treatment [17].

Despite the promise, there are several challenges with the clinical adoption of medical images for precision cervical cancer management. The correlation between images and outcomes is not always robust. For example, pretreatment ADC was not shown to be predictive of cervical cancer response to CRT [12]. In another study, FDG-PET offered no added prediction value to MR prediction [13]. The problem's complexity increases with available multiparametric MR, including perfusion, T1-, and T2-weighted information. Furthermore, PET and CT radiomics analyses can involve different feature extraction and classification methods. Unfortunately, longitudinal multimodal imaging studies with PET and MR may be cost-prohibitive as a routine clinical practice. To reduce the complexity and improve standardization of cervical cancer outcome prediction, we aim to answer three questions: 1. What imaging modality best predicts treatment response? 2. When is the optimal time point for predictive imaging studies? 3. What radiomic features and classification methods are best suited for cervical cancer outcome prediction?

## 2. Overview of the Dataset

The dataset used in this paper is obtained from the Cervical Cancer Tumor Heterogeneity (CCTH) collection in The Cancer Imaging Archive (TCIA) [18,19], including twenty-three cervical cancer patients. Functional MRIs consisting of T1-weighted and T2-weighted, dynamic contrast-enhanced (DCE), diffusion-weighted (DWI), and post-contrast MRI, as well as FDG

PET/CT, were obtained in parallel and prospectively timed with the radiation therapy course. Imaging was performed at three time-points/radiation dose levels: before treatment start (dose 0), early during the treatment course (2-2.5 weeks after treatment start/dose 20-25 Gy), and at mid-treatment (4-5 weeks after treatment start/dose 45-50 Gy). For simplicity, in this study, we denote before treatment start as 'Time-1', 2-2.5 weeks after treatment as 'Time-2', and mid-treatment as 'Time-3'. For example, 'Time-12' represents data included in both 'Time-1' and 'Time-2'. Manual contours (regions of interest) of the tumor volumes, as defined by T2-weighted MRI and co-registration with PET/CT, are included in the CCTH for each case and each imaging time point. Progression-free survival is the primary endpoint. Responders or non-responders to radiotherapy are defined by 1-month post-treatment tumor volume regression < 10% residual volume from baseline. A full table of patient characteristics is provided in Appendix A.

The dataset consists of 2D slices of different modalities and a 3D tumor mask for each patient. Since the image size can vary across different patients, we extract the medical images from the tumor region with the given mask as the region of interest (ROI) and then harmonize the matrix dimension to 256X256X32 for feature extraction.

## 3. Feature Extraction and Selection

### 3.1 Categories of imaging features

The extracted features are divided based on orders.

*Baseline Features (Zero Order Features)*

Baseline features include tumor size, max intensity and mean intensity.

*First Order Features*

First order features are also known as histogram-based features, including mean intensity, max intensity, variance, percentiles, skewness, and kurtosis. Detailed expressions of individual terms are shown in Appendix B .

*Second Order Features (Texture Features)*

Texture features involve the analysis of relationships between pairs of pixels or voxels. Texture features are usually computed with the Gray-Level Co-occurrence Matrix (GLCM). GLCM elements are defined as in Equation 1.

*Equation 1*

$$C_{\Delta x,\Delta y}(i,j) = \sum_{x=1}^{n}\sum_{y=1}^{m} \mathbb{I}_{[I(x,y)=i, I(x+\Delta x, y+\Delta y)=j]}$$

where, 'I' is the gray-level image, 'i' and 'j' are pixel values. 'n' and 'm' are the sizes of the image, (x,y) is the starting position, and (Δx, Δy) represents the offset from starting position. In this paper, we use standard Haralick texture features, including contrast, energy, entropy, and homogeneity [20,21]. The definition of each statistical term is summarized in Appendix C.

For the 3D tumor under different modalities, we consider extracting GLCM features from individual 2D slices (GLCM2D) as well as directly from the 3D geometry (GLCM3D). GLCM2D features are extracted in four different in-plane directions, while GLCM3D features are extracted in thirteen different directions. Regarding Haralick's features, all directions are considered symmetrically. Table 1 summarizes the directions for GLCM2D and GLCM3D.

*Table 1 GLCM 2D&3D Directions*

| | | | | | |
|---|---|---|---|---|---|
| **GLCM2D** | (0, 1) | (1, 0) | (1, 1) | (1, -1) | |
| **GLCM3D** | (0, 1, 0) | (1, 0, 0) | (1, 1, 0) | (1, -1, 0) | (0, 0, 1) |
| | (0, 1, 1) | (1, 0, 1) | (1, 1, 1) | (1, -1, 1) | |
| | (0, -1, 1) | (-1, 0, 1) | (-1, 1, 1) | (-1, -1, 1) | |

*Higher Order Features (Gabor Filter)*

Compared to zero, first, and second-order features, higher-order features extract more abstract features beyond the local neighborhoods. Typical higher-order features include Gabor filters, Fourier transformations, and wavelet transformations. Among different higher-order features, the Gabor filter is particularly effective in capturing texture features with frequency and direction representation [22]. Gabor filter is often employed for edge detection and texture analysis, like mammogram tumor classification and architectural distortion detection. In this paper, we compute the statistical features after applying the Gabor filter to the original medical image, including mean intensity, max intensity, variance, skewness, and kurtosis. Specifically, we consider various Gabor filter parameters, including different sizes (10-by-15, 15-by-10, 15-by-15), frequencies (0.6, 0.8), and angles (0°, 45°, 90°, 135°).

## 3.2 Feature Selection Method

Four feature selection methods widely used in medical imaging research were employed in this study [23], including Recursive Feature Elimination (RFE), Random Forest (RF), Gradient Boosting (GB), and Principal Component Analysis (PCA).

# 4. Results

The prediction performance is primarily measured by the area under the curve (AUC) with five-fold validation. Additionally, three-fold validation is employed to assess the sensitivity of prediction result to variations in the data.

## 4.1 Overall prediction accuracy

We predict the treatment responder/non-responder using the image features extracted from different methods, as discussed in Section 3.1. Table 2 summarizes the prediction accuracy of different image features extracted at varying time points. From the given results in the table, we can observe that:

1. GLCM-based features extracted from 2D slices have the best overall performance regarding prediction accuracy. Combining GLCM2D with zero order or first order features could further enhance the prediction accuracy, but the top contributing features are still from the GLCM2D based on the RFE analysis.
2. For the same feature extraction method, adding more time points, in general, increases the prediction accuracy, while different features show different sensitivity to new data and different trends to overfitting issues.
3. If a single time point is used, Time-3 performed better compared to Time-1 and Time-2. A reasonable explanation of this is that tumor response to treatment is more evident toward later time points and indicative of the patient response.
4. If two time points are used, the combinations, including Time-3 (Time-13 and Time-23) are usually better than those without it (Time-12), except for the Gabor filter. The performance of Time-13 vs. Time-23 depends on the feature selection. This can be similarly explained by tumor response to treatment.

One thing to note is that these accuracy values are obtained using all the features extracted from the medical images. A detailed model complexity analysis is shown in Appendix D.

*Table 2 Prediction accuracy at different time points with all modalities*

|  | Baseline | First Order | GLCM2D | GLCM2D Zero | GLCM2D First | GLCM3D | GLCM3D Zero | GLCM3D First | Gabor Zero | Gabor First |
|---|---|---|---|---|---|---|---|---|---|---|
| Time-1 | 0.50 | **0.63** | 0.45 | 0.42 | 0.50 | 0.43 | 0.38 | 0.53 | 0.48 | 0.58 |
| Time-2 | 0.53 | 0.42 | 0.57 | **0.60** | 0.48 | 0.45 | 0.42 | 0.38 | 0.58 | 0.58 |
| Time-3 | 0.53 | **0.72** | 0.70 | 0.68 | **0.72** | 0.50 | 0.43 | 0.68 | 0.50 | 0.63 |
| Time-12 | 0.43 | 0.38 | 0.57 | 0.62 | **0.67** | 0.42 | 0.30 | 0.55 | 0.60 | 0.60 |
| Time-13 | 0.38 | **0.72** | 0.58 | 0.63 | 0.60 | 0.50 | 0.43 | **0.72** | 0.40 | 0.45 |
| Time-23 | 0.42 | 0.62 | 0.57 | **0.70** | 0.60 | 0.50 | 0.47 | 0.62 | 0.52 | 0.40 |
| Time-123 | 0.43 | 0.55 | 0.73 | **0.75** | 0.70 | 0.50 | 0.45 | 0.55 | 0.47 | 0.40 |

## 4.2 Prediction with fewer modalities

We further extend our study to understand the prediction accuracy using a single modality, including PET/CT (denoted as PET), DCE, and DWI-ADC (denoted as ADC), or the combinations of any two modalities. The prediction accuracy using all three time points is shown in Table 3 with five-fold validation. The results show that GLCM2D-based features have better prediction accuracy than other feature extraction methods, and combining GLCM2D with zero order features has slightly better performance than combining with first order features. This conclusion is consistent with the result when using all modality images shown in Section 4.1. Moreover, we see an overall advantage of using ADC for single modality prediction and DCE&PET for dual modalities prediction. An exception is observed with DCE-based Gabor Zero features, which initially yields the highest prediction accuracy in five-fold validation, though this result is inconsistent with other data in the same column. To verify this anomaly, we conduct an additional three-fold validation, as shown in Table 4. The new results indicate that DCE-based Gabor Zero features no longer produce the highest accuracy, while other features that perform well in the five-fold validation remained reliable. Therefore, we conclude that GLCM2D-based features show greater stability, and are generally superior to other feature types, as detailed in Section 5.

## 4.3 Understanding the prediction power of GLCM features

We further study feature selection to reduce model complexity and understand the significance of individual features using RFE, PCA, GB, and RF. Figure 1 shows AUC performance with varying features extracted from ADC of all three time points. Here, we consider feature numbers from 3 to 20 due to limited data size. The results show relatively stable performance with reducing features to three.

To understand the prediction power of individual GLCM features, we chose the ADC GLCM feature, randomly selected the top 5, 12, and 20 features, and summarized the names of features in Table 5. Among different types of GLCM features, contrast is the most significant feature, and correlation also plays an important part in certain cases, while energy and homogeneity are not significant parameters in determining treatment effectiveness.

*Table 3 Prediction accuracy with fewer modalities using all time points (k=5)*

|  | Baseline | First Order | GLCM2D | GLCM2D Zero | GLCM2D First | GLCM 3D | GLCM3D Zero | GLCM3D First | Gabor Zero | Gabor First |
|---|---|---|---|---|---|---|---|---|---|---|
| ADC | 0.58 | 0.53 | **0.75** | **0.75** | 0.68 | 0.60 | 0.50 | 0.43 | 0.48 | 0.67 |
| DCE | 0.43 | 0.37 | 0.67 | 0.60 | 0.57 | 0.60 | 0.47 | 0.37 | **0.80** | 0.50 |
| PET | 0.63 | **0.72** | 0.65 | 0.62 | 0.62 | 0.57 | 0.63 | **0.72** | 0.43 | 0.27 |
| ADC&DCE | 0.40 | 0.40 | **0.75** | 0.60 | 0.60 | 0.50 | 0.37 | 0.40 | 0.50 | 0.72 |
| ADC&PET | 0.60 | 0.57 | **0.68** | **0.68** | 0.60 | 0.50 | 0.57 | 0.57 | 0.42 | 0.47 |
| DCE&PET | 0.62 | 0.57 | 0.68 | **0.77** | 0.75 | 0.65 | 0.63 | 0.60 | 0.50 | 0.38 |
| ADC&DCE&PET | 0.43 | 0.55 | 0.73 | **0.75** | 0.70 | 0.50 | 0.45 | 0.55 | 0.47 | 0.40 |

*Table 4 Prediction accuracy with fewer modalities using all time points (k=3)*

|  | Baseline | First Order | GLCM2D | GLCM2D Zero | GLCM2D First | GLCM 3D | GLCM3D Zero | GLCM3D First | Gabor Zero | Gabor First |
|---|---|---|---|---|---|---|---|---|---|---|
| ADC | 0.58 | 0.42 | 0.67 | **0.71** | 0.65 | 0.50 | 0.43 | 0.31 | 0.53 | 0.5 |
| DCE | 0.5 | 0.53 | 0.63 | **0.64** | 0.57 | 0.61 | 0.50 | 0.53 | 0.63 | 0.49 |
| PET | 0.43 | **0.69** | 0.64 | 0.64 | 0.60 | 0.50 | 0.47 | **0.69** | 0.43 | 0.36 |
| ADC&DCE | 0.44 | 0.49 | **0.71** | 0.60 | 0.64 | 0.50 | 0.44 | 0.43 | 0.64 | 0.58 |
| ADC&PET | 0.56 | 0.60 | **0.75** | 0.69 | 0.65 | 0.50 | 0.51 | 0.54 | 0.49 | 0.57 |
| DCE&PET | 0.53 | 0.64 | 0.75 | 0.75 | **0.79** | 0.61 | 0.58 | 0.64 | 0.53 | 0.47 |
| ADC&DCE&PET | 0.49 | 0.69 | 0.68 | **0.74** | **0.74** | 0.50 | 0.49 | 0.64 | 0.64 | 0.51 |

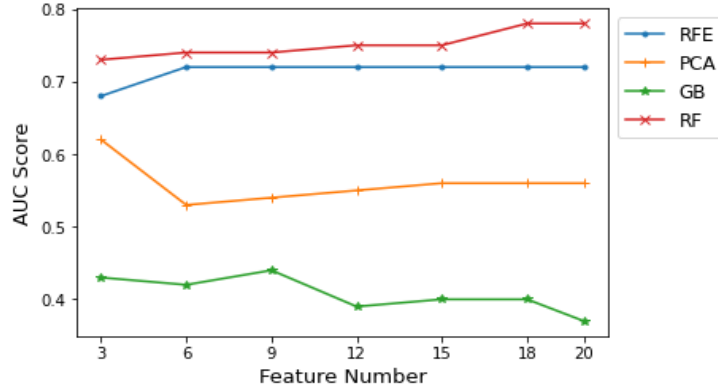

*Figure 1 Prediction accuracy from different feature extraction methods*

*Table 5 Example of GLCM feature selection from individual modality (CON refers to Contrast, COR refers to correlation, ENE refers to Energy, HOM refers to Homogeneity)*

|  |  | ADC | | | | DCE | | | | PET | | | |
|---|---|---|---|---|---|---|---|---|---|---|---|---|---|
|  | # Features | CON | COR | ENE | HOM | CON | COR | ENE | HOM | CON | COR | ENE | HOM |
| Time-1 | 5 | 4 | 1 | 0 | 0 | 5 | 0 | 0 | 0 | 5 | 0 | 0 | 0 |
|  | 12 | 8 | 4 | 0 | 0 | 10 | 2 | 0 | 0 | 10 | 2 | 0 | 0 |
|  | 20 | 12 | 8 | 0 | 0 | 13 | 7 | 0 | 0 | 14 | 6 | 0 | 0 |
| Time-12 | 5 | 4 | 1 | 0 | 0 | 5 | 0 | 0 | 0 | 5 | 0 | 0 | 0 |
|  | 12 | 9 | 3 | 0 | 0 | 12 | 0 | 0 | 0 | 11 | 1 | 0 | 0 |
|  | 20 | 14 | 6 | 0 | 0 | 17 | 3 | 0 | 0 | 15 | 5 | 0 | 0 |
| Time-123 | 5 | 4 | 1 | 0 | 0 | 5 | 0 | 0 | 0 | 5 | 0 | 0 | 0 |
|  | 12 | 11 | 1 | 0 | 0 | 12 | 0 | 0 | 0 | 11 | 1 | 0 | 0 |
|  | 20 | 16 | 4 | 0 | 0 | 18 | 2 | 0 | 0 | 18 | 2 | 0 | 0 |

## 5. Discussions

Here, we studied image-based outcome prediction for cervical cancer patients undergoing CRT. Our paper has explored several vital questions, including the prediction effectiveness of different modalities and modalities measured at different treatment stages.

*Prediction results with different imaging modalities*

Previous studies explored either the prediction with a single modality [3, 4, 7] or multiple modalities [9, 17]. There has not been a consensus as to which imaging modality is best for cervical cancer outcome prediction [24]; Our analysis has shown that for the three individual modalities considered in cervical cancer treatment response prediction, MRI-based ADC has the highest prediction accuracy, followed by PET/CT, while MRI-based DCE has the lowest prediction accuracy. When considering dual modalities, the combination of MRI-based DCE and PET/CT has the highest prediction accuracy.

Moreover, our GLCM-based method shows that Contrast and Correlation are the two most important factors in predicting the treatment response, which is consistent with the fact that MRI provides superior soft tissue contrast [25].

This paper found that ADC better predicts than DCE for cervical cancer treatment response, indicating the significance of cellularity in tumor aggressiveness and response to therapy. This observation is consistent with previous reports [26,27]. Though DCE has been reported to enhance detectability of residual cancer [28,29], there is not existing findings validating the superior of DCE Gabor filter features over the texture features. When utilizing dual modalities, we found the combination of DCE and PET to be slightly superior for response prediction [31] due to their complementary information [30].

## Prediction results with fewer time points

Considering a single time point, modality images measured at Time-3 have better performance over other individual time points, including the combination of time points 1 and 2. This is not surprising as Time-3 reflects the post-treatment tumor status. A similar conclusion is reported in laryngeal carcinoma evaluation [32]. Unfortunately, imaging information at this time point cannot be used to adapt the treatment plan that has already been delivered. When adding more time points, the prediction accuracy generally increases compared to a single time point. Adding more time points helps establish the temporal dynamics of the tumor responding to treatment [33]. However, in this study, compared to using all three time points, the prediction accuracy of Time-3 only decreases by around 4% with GLCM2D features and decreases by around 9% with GLCM2D with zero order features, indicating minimal contribution to the prediction with the temporal information.

## Feature extraction methods

Compared to the previous study [17], which uses the first-order features, our paper studies the effectiveness of different feature extraction methods and shows that for the cervical cancer data considered, the GLCM features hold a clear advantage, which we attribute to the GLCM features' robustness. Since GLCM considers the relative spatial relationship between pixels, it tends to be less sensitive to variation in image acquisition parameters among patients and image modalities. Although adding zero and first-order features could be complementary to texture features, considering single modality and multiple modalities, texture features always play a dominant role in determining treatment effectiveness. This observation is consistent with other radiotherapy research, such as in lung tumors, bladder tumors, and glioblastoma [34, 35, 36, 37]. Moreover, for individual GLCM features extracted from the images, we show that Contrast is the most predictive GLCM feature, while Energy and Homogeneity are not significant in determining the treatment response. The results show that the contrast of the tumor is more indicative of its response to CRT than its energy and homogeneity. In comparison, the uniformity and concentration of pixels reflected by Energy and Homogeneity are relatively insensitive to reflecting the underlying tumor aggressiveness and response, which is consistent with pulmonary tumor analysis [38,39,40].

*Prediction methods*

Besides the comparison of feature extraction methods, another contribution of this paper is the comparison of feature selection methods. In this study, we show that the RFE and RF are more effective than other feature extraction methods. RFE has been proven to be an efficient way to select features in different tumor research, like tumor classifications and predictions in cervical cancer [17], prostate cancer [41], and lung cancer [42]. Similarly, RF has been successfully applied to different types of medical image classification, labeling, and segmentation [43].

RFE-based approaches have demonstrated superiority, particularly in medical applications where the number of features far exceeds the number of samples [44,45]. Unlike PCA, which transforms original features into a new set of uncorrelated components, or decision-tree-based methods like RF and GB, RFE focuses on selecting the most critical features for prediction, significantly simplifying the feature extraction process. However, RFE's focus on feature selection can make it more susceptible to overfitting, especially with smaller datasets or when the feature selection process is overly aggressive. In contrast, decision-tree-based methods like RF and GB are more complex but offer greater robustness against overfitting. In this study, RF slightly outperformed RFE, though their relative performance may vary depending on the specific dataset.

## 6. Conclusions

This paper explores how multi-modality images (PET/CT, ADC, DCE) at different treatment points can be more effectively used to predict treatment response for cervical cancer. Our research explores different image feature extraction methods and feature selection methods to understand the best approach for predicting cervical cancer treatment response.

The key conclusions and contributions of this paper are:

1. We discover the GLCM-based features are more effective compared to other types of image features considered regarding prediction accuracy. Moreover, for GLCM-based features extracted from 2D and 3D spaces, 2D GLCM features are more effective in predicting the treatment response. We further discover that within the GLCM features, Contrast and Correlation are the main contributors to the prediction from RFE extraction.
2. We compare the prediction effectiveness between different treatment points. Our results have shown that post-stage (Time-3) measurement has the highest prediction accuracy than pre-stage (Time-1) and mid-stage (Time-2). When using GLCM features, post-stage measurement is only 4% lower than using all time points with 2D GLCM features and 9% lower than using 2D GLCM with zero order features, indicating minimal contribution to the prediction with the temporal information.
3. We further compare the prediction accuracy between different modalities and their combinations. We discover that ADC has the best prediction performance compared to PET/CT, followed by DCE with GLCM-based features. Also, combining DCE and PET/CT is more effective compared to the other two possible combinations. These results not only provide a comprehensive comparison of the relationship between different modalities and time points, but also indicate the potential to reduce the measurements and enhance cervical cancer treatment efficiency.

# Supplementary Document

## A. Patient Characteristics Table

Table A.1 Patient clinical data

| Patient ID | Time Measurement | Early Response | Local Recurrence | Distant Metastasis | Death from Cervical Cancer |
|---|---|---|---|---|---|
| 1 | Time-1, Time-2, Time-3 | Non-responder | y | n | y |
| 2 | Time-1, Time-2, Time-3 | Non-responder | y | n | y |
| 3 | Time-1, Time-2, Time-3 | Responder | n | n | n |
| 4 | Time-1, Time-2, Time-3 | Non-responder | n | y | y |
| 5 | Time-1, Time-2, Time-3 | Responder | n | n | n |
| 6 | Time-1, Time-2, Time-3 | Responder | n | n | n |
| 7 | Time-1, Time-2, Time-3 | Non-responder | y | y | y |
| 8 | Time-1, Time-2, Time-3 | Non-responder | y | y | y |
| 9 | Time-1, Time-2, Time-3 | Responder | n | n | n |
| 10 | Time-1, Time-2, Time-3 | Responder | n | n | n |
| 11 | Time-1, Time-2, Time-3 | Responder | N/A | y | y |
| 12 | Time-1, Time-2, Time-3 | Responder | n | n | n |
| 13 | Time-1, Time-2, Time-3 | Responder | n | n | n |
| 14 | Time-1, Time-2, Time-3 | Non-responder | n | n | n |
| 15 | Time-1, Time-2, Time-3 | Responder | n | n | n |
| 16 | Time-1, Time-2, Time-3 | Responder | n | n | n |
| 17 | Time-1, Time-2 | Responder | n | n | n |
| 18 | Time-1, Time-2, Time-3 | Non-responder | n | y | y |
| 19 | Time-1, Time-2, Time-3 | Non-responder | y | y | y |
| 20 | Time-1, Time-2, Time-3 | Responder | n | n | n |
| 21 | Time-1, Time-2, Time-3 | Responder | n | n | n |
| 22 | Time-1, Time-2, Time-3 | Non-responder | n | n | n |
| 23 | Time-1, Time-2, Time-3 | Non-responder | n | n | n |

## B. First Order feature expressions

From the histogram of the medical image, denote the probability of occurrence of an intensity value k as p(k). Then each statistical term can be described as:

$$Mean = \sum_{k} k \times p(k)$$

$$Max = \max(k)$$

$$Variance = \sum_{k}(k - Mean)^2 \times p(k)$$

$$Skewness = \frac{\sum_{k}(k - Mean)^3 \times p(k)}{Variance^{3/2}}$$

$$Kurtosis = \frac{\sum_{k}(k - Mean)^4 \times p(k)}{Variance^2}$$

## C. GLCM texture feature expressions

This section provides detailed definitions of each statistical feature from GLCM, denoted by matrix m.

$$Contrast = \sum_{i}\sum_{j}(i - j)^2 m(i,j)$$

$$Correlation = \frac{\sum_{i}\sum_{j} ij[m(i,j) - \mu_x\mu_y]}{\sigma_x\sigma_y}$$

$$Energy = \sum_{i}\sum_{j} m(i,j)^2$$

$$Homogeneity = \sum_{i}\sum_{j}\frac{m(i,j)}{1 + |i - j|}$$

## D. Modal complexity and overfitting analysis

To understand how different numbers of features influence the model's prediction performance, we use the RFE to pick the features delivering the best prediction accuracy computed with AUC score. As an example, we compare the prediction accuracy using 2D GLCM features extracted from ADC using zero order, first order and second order (GLCM2D) features. The prediction accuracy is shown in Figure D.1. From the figures, we notice that the prediction accuracy converges when using the top 20-30 features, and the training accuracy is higher than testing accuracy, indicating the existence of potential overfitting.

To understand the potential overfitting, we further quantify the balance between fit and complexity with the Akaike Information Criterion (AIC) value. The profile of AIC values with different types and different numbers of image features is shown in Figure D.2. Interestingly, although the GLCM-based features have a better prediction, the AIC curve is nearly linear to the number of features, indicating for the GLCM-based prediction model, adding additional complexity to the model is unnecessary and a feature selection method would be desired.

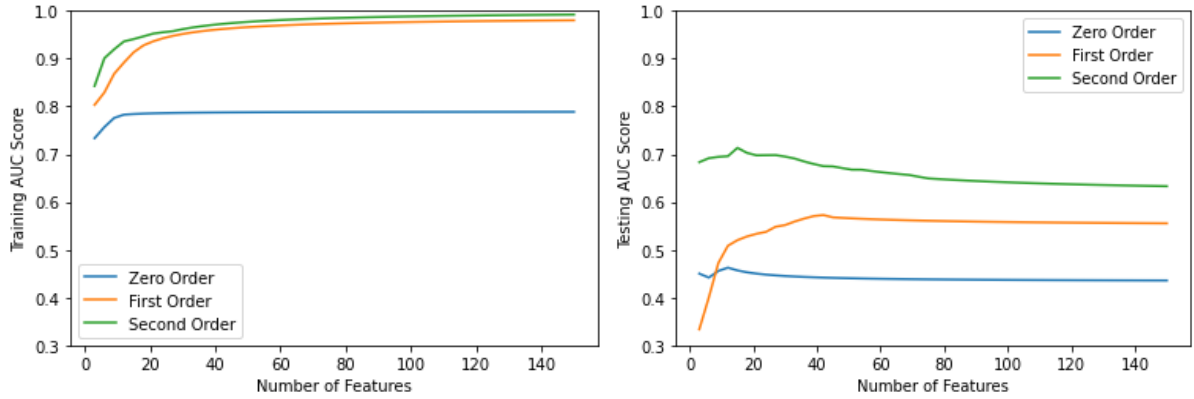

*Figure D.1 AUC score for different number of image features*

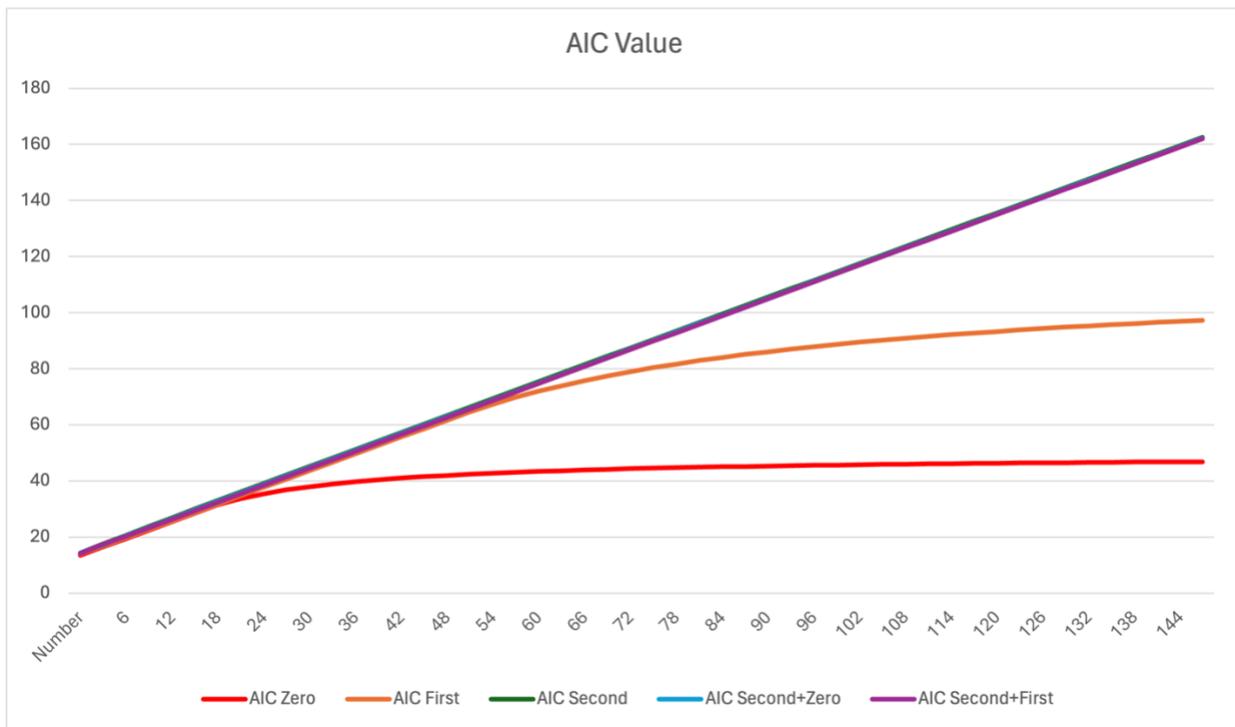

*Figure D.2 AIC value for different feature numbers ('Zero' means baseline zero order features, 'First' means first order features, 'Second' means second order GLCM2D features)*